# Infrared ellipsometry study of the photo-generated charge carriers at the (001) and (110) surfaces of SrTiO$_3$ crystals and the interface of corresponding LaAlO$_3$/SrTiO$_3$ heterostructures

M. Yazdi-Rizi[1,†], P. Marsik[1], B.P.P. Mallett[1,2], K. Sen[1], A. Cerreta[1], A. Dubroka[3], M. Scigaj[4], F. Sánchez[4], G. Herranz[4] and C. Bernhard[1,‡]

[1]Physics Department and Fribourg Center for Nanomaterials (FriMat), University of Fribourg, Chemin du Musée 3, CH-1700 Fribourg, Switzerland
[2]Robinson Research Institute, Victoria University, P.O. Box 600, Wellington, New Zealand
[3]Department of Condensed Matter Physics, Faculty of Science and Central European Institute of Technology, Masaryk University, Kotlářská 2, 61137 Brno, Czech Republic
[4]Institut de Ciència de Materials de Barcelona (ICMAB-CSIC), Campus de la UAB, Bellaterra 08193, Catalonia, Spain

With infrared (IR) ellipsometry and DC resistance measurements we investigated the photo-doping at the (001) and (110) surfaces of SrTiO$_3$ (STO) single crystals and at the corresponding interfaces of LaAlO$_3$/SrTiO$_3$ (LAO/STO) heterostructures. In the bare STO crystals we find that the photo-generated charge carriers, which accumulate near the (001) surface, have a similar depth profile and sheet carrier concentration as the confined electrons that were previously observed in LAO/STO (001) heterostructures. A large fraction of these photo-generated charge carriers persist at low temperature at the STO (001) surface even after the UV light has been switched off again. These persistent charge carriers seem to originate from oxygen vacancies that are trapped at the structural domain boundaries which develop below the so-called antiferrodistortive transition at $T^* = 105$ K. This is most evident from a corresponding photo-doping study of the DC transport in STO (110) crystals for which the concentration of these domain boundaries can be modified by applying a weak uniaxial stress. The oxygen vacancies and their trapping by defects are also the source of the electrons that are confined to the interface of LAO/STO (110) heterostructures which likely do not have a polar discontinuity as in LAO/STO (001). In the former, the trapping and clustering of the oxygen vacancies also has a strong influence on the anisotropy of the charge carrier mobility. We show that this anisotropy can be readily varied and even inverted by various means, such as a gentle thermal treatment, UV irradiation, or even a weak uniaxial stress. Our experiments suggest that extended defects, which develop over long time periods (of weeks to months), can strongly influence the response of the confined charge carriers at the LAO/STO (110) interface.

## I) INTRODUCTION

The discovery of a nearly two-dimensional electron gas (2DEG) with a high mobility at the interface of LAO/STO heterostructures [1], and the demonstration of subsequent electric-field-effect devices [2, 3] have renewed the interest in the extraordinary structural and electronic properties of STO. At room temperature, it seems to be a rather ordinary material with a simple perovskite structure with cubic symmetry [4]. It is also a band-insulator with an indirect band-gap of 3.25 eV and a direct one of 3.75 eV [5]. Thanks to the flexible valence of the Ti ions, which can be varied continuously between 4+ and 3+, one can readily dope the conduction band of STO (it has mainly Ti-3$d$ $t_{2g}$ character) with electrons to make it metallic and even superconducting at very low temperature, for example by a partial replacement of Sr$^{2+}$ with La$^{3+}$, Ti$^{4+}$ with Nb$^{5+}$ or by creating oxygen vacancies [6-9].

STO undergoes several structural phase transitions as the temperature is lowered. The first one is the so-called antiferrodistortive (AFD) transition at $T^* = 105$ K from the cubic to a tetragonal symmetry. It involves an antiphase rotation of the neighboring TiO$_6$ octahedra around the c-axis and leads to a doubling of the unit cell along all three directions and a slight mismatch of the lattice parameters along the rotation axis (the c-axis) and the perpendicular direction (a-axis) with c/a ≈ 1.0015 at low temperature [10]. Unless special care is taken, e.g. by applying a uniaxial pressure [11-13] or a large electric field [14], this results in a poly-domain state with strained boundaries between the domains with different c-axis orientation. Further structural transitions have been reported to an orthorhombic state at 65 K and a rhombohedral one at 37 K [15]. In this low temperature range STO also becomes a so-called "incipient ferroelectric" or "quantum paraelectric" material for which a ferroelectric transition is only prohibited by the quantum fluctuations of the lattice [4]. This is accompanied by a dramatic softening of a corresponding IR-active phonon mode which involves the off-center displacement of the Ti ion with respect to the surrounding oxygen octahedron [4]. The eigenfrequency of this so-called "soft-mode" undergoes a rather dramatic decrease from about 95 cm$^{-1}$ at room temperature to about 7 cm$^{-1}$ and 16 cm$^{-1}$ at 10 K along the a- and c-axis, respectively [16], and the dielectric constant of STO exhibits a corresponding increase from $\varepsilon \approx 300$ at room temperature to $\varepsilon > 20000$ at 10 K [17,18].

To induce the ferroelectric transition it is already sufficient to weaken the quantum fluctuations by replacing the lighter oxygen isotope $^{16}$O with the heavier $^{18}$O [19]. A ferroelectric order can also be readily induced with strain effects and various kinds of defects [20]. For example, room temperature ferroelectricity has been observed in STO thin films that are grown under tensile strain on top of a DyScO$_3$ substrate [20]. Moreover, a polar order that is likely induced by a flexoelectric effect has been observed in the vicinity of the domain boundaries of the structural poly-domain state that appears below $T^*$ [21-25] and can even modify the magnetic state of a magnetic layer on STO [26]. It has even been reported that these polar domain boundaries are conducting, likely because of trapped oxygen vacancies [27]. These examples demonstrate that the electronic properties of STO are extremely versatile and very susceptible to the local structure, defects and strain.

It is probably for similar reasons that the origin of the confined electrons at the LAO/STO interface is still debated. The explanations range from (i) a discontinuity in the stacking sequence of the polar atomic layers that leads to a divergence of the electrostatic potential [1, 2] (a so-called "polar catastrophe") that is compensated by an electron transfer from the surface of the LAO layer to the one of STO, (ii) an inter-diffusion of cations like $La^{3+}$ and $Sr^{2+}$ [28], (iii) oxygen vacancies that are trapped near the interface [29] or (iv) a mixture of the effects (i)-(iii). For the LAO/STO (001) interface it has been shown in favor of scenario (i) that the interface termination with a LaO layer on top of a $TiO_2$ layer is a crucial factor and that a threshold thickness of the LAO layer of at least 4 unit cells is required to obtain the 2DEG [2]. It was also shown that a certain carrier concentration is maintained even after an extensive post-growth annealing of the samples in oxygen rich atmosphere. Interestingly, 2DEGs have been also found at interfaces oriented along (110) and (111) [30, 31]. Much the same as in (001) interfaces, the conduction at (110) and (111) interfaces appeared also above a critical thickness in samples that underwent an extensive post-growth annealing in oxygen rich atmosphere. The (110) interface is particularly interesting as, nominally, no polar discontinuity is present. In this latter, whether the origin of its conductance is caused by effects other than polar discontinuity has remained so far unclear. Finally, there is wide consensus that the conductance in heterostructures of amorphous LAO on STO (001) and of γ-$Al_2O_3$ on $SrTiO_3$ (001), of which the latter yields record values of the low-T mobility [32], has its origin in the formation of oxygen vacancies at the interface.

The oxygen vacancies are also playing an important role in the photo-generation of mobile charge carriers via the absorption of UV light [33]. It was earlier recognized that a strong photo-doping effect can be achieved even in bare STO if the photon energy exceeds the band-gap of STO of about 3.2 eV. Especially puzzling is the circumstance that these photo-generated charge carriers can persist for a long time after the UV-radiation has been turned off [34, 35]. It has been suspected that oxygen vacancies are playing an important role in this persistent photo-effect, but there is still no consensus about the mechanism of the creation of these oxygen vacancies, their electronic structure or their interaction with defects and with each other. Recent angle resolved photo-emission spectroscopy (ARPES) studies have confirmed that the UV-light induces oxygen vacancies and, at the same time, generates a 2DEG at the bare surface of STO [36-38].

The confinement of the electrons was shown to give rise to a series of mini-bands. A polarization analysis also distinguished the bands with $d_{xy}$ symmetry from the ones with $d_{xz/yz}$ symmetry (the latter are slightly higher in energy). A corresponding photo-induced or photo-enhanced conductivity was subsequently observed in the LAO/STO heterostructures [39-41].

Theoretical calculations have confirmed that oxygen vacancies can be stabilized near the surface of STO [42] and may further reduce their energy by forming bi- or tri-vacancies [22, 43] or even larger clusters that are preferentially located in the vicinity of structural domain boundaries [44]. Calculations that include a strong on-site Coloumb-repulsion between the two electrons that are left behind when atomic oxygen leaves the sample suggest that only one of the electrons is trapped by the oxygen vacancy whereas the second one may remain delocalized [45]. Such singly-charged oxygen vacancies are also suspected to cause the magnetic phenomena which have been observed in some of the LAO/STO heterostructures [45-47]. This hypothesis is supported by a recent circular dichroism study which found optically-induced, persistent magnetism in oxygen deficient STO [48]. Nevertheless, it remains to be further explored under which conditions such singly charged oxygen vacancies remain stable and thus lead to persistent conductivity (and possibly also magnetism).

Here we present a combined IR ellipsometry and DC transport study of the photo-induced charge carriers in single crystals of STO (001) and STO (110) and heterostructures of LAO/STO (001) and LAO/STO (110). In section II we describe first the sample growth and preparation and the experimental techniques that have been used in this work. The studies of the photo-doping effect of the bare STO (001) and STO (110) single crystals are presented in sections III.A and III.B, respectively. The corresponding IR-experiments on LAO/STO (001) and LAO/STO (110) heterostructures are discussed in sections III.C and D. A brief summary is given in section IV.

## II) EXPERIMENTS

### II.A.) Sample preparation and growth

The LAO/STO (001) heterostructure used in this work is identical to the sample that was measured in Ref. [49]. The LAO/STO (110) sample with a thickness of 10 LAO monolayers has been grown as reported in Ref. [30]. Namely, the LAO (110) thin film was grown by pulsed laser deposition (PLD) at an oxygen pressure of $10^{-4}$ mbar at a laser repetition rate of 1 Hz and a pulse energy of about 26 mJ. At the end of the deposition, the sample was cooled down in oxygen rich atmosphere to minimize the formation of oxygen vacancies under a pressure of 0.3 mbar from T = 850 °C to 750 °C and 200 mbar from T = 750 °C to room temperature, with a dwell time of 1 hour at 600 °C.

The bare STO (110) and STO (001) single crystals were purchased from SurfaceNet and had dimensions of 10×1×1 $mm^3$. The STO (110) substrate for the LAO/STO (110) heterostructures was purchased from CrysTec GmbH.

The UV-light used to photo-dope the samples is from a 100 W Xe-lamp that covers a broad range of wavelengths from about 200 to 1000 nm. With a silver coated concave mirror, the light from this lamp was focused on the surface of the sample that was mounted in a He-flow cryostat. The focal spot had a size of about 1.5 cm$^2$ and covered the sample almost homogeneously.

### II.B.) DC transport

The DC resistance of the samples was measured as a function of time and temperature with a KEITHLEY 2600 current source using a four-probe method while the sample was mounted in the cryostat of the IR ellipsometer. For the electrical contacts we glued gold wires with silver paste to the sample surface. The current was set to 1 μA while the voltage was recorded. The upper limit of the voltage that could be measured was 40 V.

### II.C.) IR spectroscopy

The IR ellipsometry measurements have been performed with a home-built setup that is equipped with a He-flow cryostat and attached to a Bruker 113V Fast-Fourier spectrometer as described in Ref. [50]. The data have been taken at different temperatures in rotating analyzer mode. The angle of incidence of the light was set to 75°. For the dark measurements without UV-illumination, special care was taken to avoid photo-doping effects by shielding the sample against visible and UV light [51].

### II.D.) Interpretation and modelling of the Berreman-mode and the dip feature

The IR ellipsometry spectra contain valuable information about the properties of the electrons that are confined at the surface of STO and the interface of LAO/STO. In this work we analyse a Berreman mode that yields direct information about the carrier concentration including its depth profile. The data are conveniently presented in terms of difference spectra of the ellipsometric angle, $\Psi = \arctan(r_p/r_s)$. The photo-doping effect on the STO surface has been measured with respect to the dark state prior to the UV illumination. For the LAO/STO (001) and (110) heterostructures we used a corresponding bare STO substrate as reference for the difference spectrum, $\Delta\Psi = \Psi(\text{sample}) - \Psi(\text{STO})$.

The spectrum of $\Delta\Psi$ contains a so-called dip-peak feature (see Fig. 1). The peak corresponds to a so-called Berreman-mode which is a plasmonic resonance where the charge carriers oscillate between the two interfaces of a conducting layer that is surrounded by insulating materials. It occurs at the screened plasma frequency of the layer, i.e. at the frequency where $\varepsilon_1(\omega_{LO}) = 0$ [49]. In case of a doped STO layer this plasma frequency can be written as,

$$\omega_{LO}^2 = \omega_{LO}^2(\text{STO}) + \omega_{pl}^2, \qquad (1)$$

where $\omega_{LO}(\text{STO}) \approx 788$ cm$^{-1}$ is the frequency of the highest longitudinal optical (LO) phonon mode of the insulating STO [52-54] and $\omega_{pl}$ the screened plasma frequency of the itinerant electrons. Consequently, the difference in frequency between the Berreman mode and $\omega_{LO}(\text{STO})$ is a measure of $\omega_{pl}$ [49, 55] that is related to the carrier concentration, $n$, like:

$$\omega_{pl}^2 = \frac{nq^2}{\varepsilon_0 \varepsilon_\infty m^*}, \qquad (2)$$

where $q$ is the electron charge, $\varepsilon_0$ the permitivity of vacuum, $\varepsilon_\infty = 5.1$ is the background dielectric constant and $m^*$ the effective mass of the itinerant electrons. The intensity of this peak is determined by the overall sheet carrier density, $N_s$, and the mobility, $\mu$ (or the inverse scattering rate, $1/\tau$).

The dip feature is a fairly sharp and pronounced minimum that occurs at a frequency of about $\omega^{\text{dip}} \approx 867$ cm$^{-1}$ where the real part of the dielectric function of the substrate of the layer, in this case STO, equals unity and thus matches the value in vacuum, $\varepsilon_1^{\text{STO}}(\omega^{\text{dip}}) = 1$. This dip feature contains contributions from the out-of-plane and the in-plane components of the dielectric function [55, 56]. The in-plane contribution arises from the Drude-like response of the itinerant electrons which leads to a reduction of $\varepsilon_1(\omega^{dip})$ and, given a sufficiently low mobility and thus large scattering rate, an increase of $\varepsilon_2(\omega^{dip})$. The strength of this dip is therefore a measure of $N_s$, but it is fairly insensitive to the details of the depth distribution of the itinerant carriers, since the penetration depth of the IR light is on the order of several micro-meters. It is also not very sensitive to the in-plane mobility of the carriers, $\mu_{ab}$, unless it is very low such that the Drude-peak in $\varepsilon_2$ is very broad and extends to the dip feature.

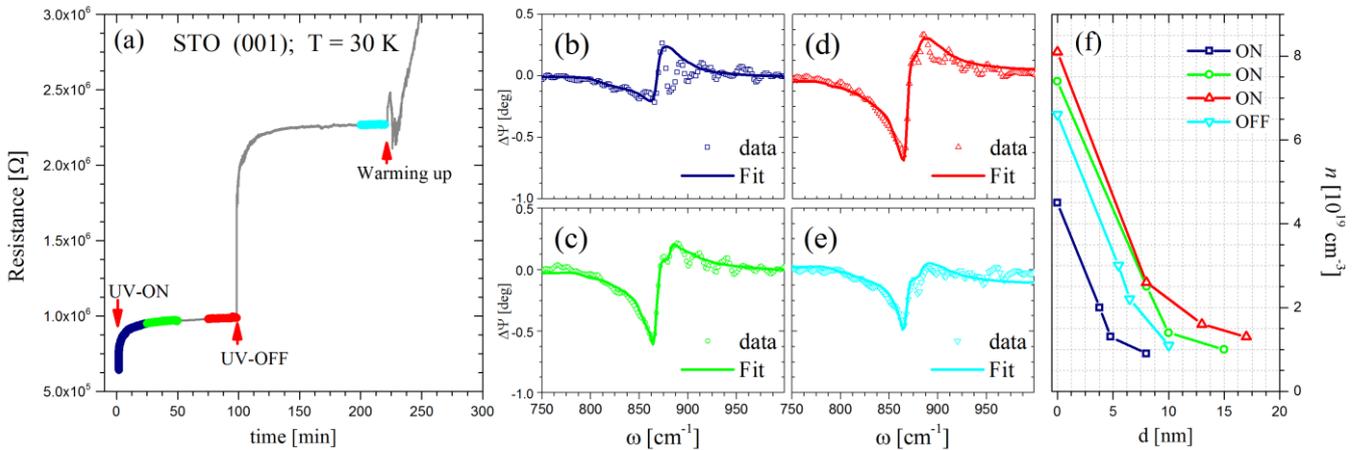

**Figure 1**: DC resistance and IR spectroscopy study of the photo-induced charge carriers at the surface of a STO (001) single crystal. **(a)** Evolution of the DC resistance during and after the illumination with UV light. Shown in color are the periods during which the ellipsometry spectra in (b)-(e) have been recorded. **(b)-(e)** Difference spectrum of the ellipsometric angle, $\Psi$, as measured at different times during and after the UV-illumination (at $t > 0$, as indicated in color in Fig. 1(a)) and in the initial, dark state at $t < 0$, i.e. $\Delta\Psi = \Psi(t > 0) - \Psi(t < 0)$. The solid lines show the best fits with the model of a conducting surface layer with a graded depth profile of the charge carriers. **(f)** Depth-profiles of the charge carrier concentration obtained from the best fits in (b)-(e). The other fit parameters are listed in Table I.

TABLE I. Parameters of the best fits to the ellipsometry data in Figs. 1(b)-1(e).

| STO (001) | $N_s$ [$10^{13}$ cm$^{-2}$] | $d$ [nm] | $\mu_{ab}$ [cm$^2$/Vs] | $\mu_\perp$ [cm$^2$/Vs] |
|---|---|---|---|---|
| Light ON; t = 45 min; (Fig. 1b) | 1.7±0.4 | 8±2 | 10±3 | 4±2 |
| Light ON; t = 90 min; (Fig. 1c) | 5.0±0.3 | 15±2 | 11±3 | 5±2 |
| Light ON; t = 180 min; (Fig. 1d) | 5.9±0.4 | 17±3 | 10±4 | 5±3 |
| Light OFF; t = 360 min; (Fig. 1e) | 3.5±0.4 | 10±3 | 8±3 | 4±3 |

For completeness, we note that for a charge carrier concentration below about $2.8 \times 10^{19}$ cm$^{-3}$ the value of $\omega_{LO}$ is smaller than $\omega^{dip}$ such that the Berreman mode is shifted to the Reststrahlen-band of STO. In that case the Berreman mode shows up in the differential spectrum $\Delta\Psi$ as a dip at $\omega_{LO}$ and the feature at $\omega^{dip}$ becomes a peak.

Throughout the paper we use, for reasons of consistency, the same value of the effective mass of $m^* = 3.2\, m_e$ ($m_e$ is the free electron mass) that was used in our earlier work [49]. We neglect here that the effective mass in STO is temperature and frequency dependent and may be different for STO (001) and STO (110). According to eq. (2) this affects the charge carrier concentration which for the case of $m^* = 1.8$, as reported for low temperatures and $\omega \approx 800$ cm$^{-1}$ in Ref. [57], would be reduced by a factor of about $1.8/3.2 \approx 0.6$.

### III) RESULTS AND DISCUSSION
### III.A.) Photo doping of STO (001)

Figure 1 shows the result of a combined DC-transport and IR-spectroscopy study of the photo-doping effect at the (001) surface of a STO single crystal. Figure 1(a) displays the time evolution of the DC-resistance at 30 K as the UV-light is switched on at t = 0 and turned-off again after a duration of about 100 minutes. The UV-light induces an immediate transition from a highly insulating state (its resistance is too high to be measured with our setup) to a conducting one with a resistance of about 600-800 kΩ and a relatively weak time dependence. From these data it is not quite clear whether the initial increase of R is an intrinsic effect or may rather be caused by the heating of the sample surface due to the UV radiation. This issue will be further addressed below in the discussion of the data in Fig. 3. The most important result of Fig. 1(a) is that the sample remains conducting even after the UV-light has been turned off. The resistance increases almost instantaneously to about 2 MΩ but then saturates and remains around 2.25 MΩ until the sample is heated and the experiment is terminated. This shows that a significant fraction of the photo-generated carriers are very long-lived with a recombination time of more than several hours.

Additional information about the evolution of the concentration and the dynamics of the photo-generated charge carriers has been obtained from the IR-ellipsometry data as shown in Figs. 1(b)-1(e). They have been recorded in parallel with the DC-transport at times that are indicated by the color scheme in Fig. 1(a). Shown are the difference spectra (with respect to the state before the UV illumination) of the ellipsometric angle, $\Delta\Psi$, in the vicinity of the highest LO-phonon mode of STO. They contain a relatively sharp dip feature around 866 cm$^{-1}$ and a weaker and broader maximum at somewhat higher frequency due to the so-called Berreman-mode. The same plasmonic feature was previously observed in LAO/STO heterostructures and was shown to contain valuable information about $N_s$ and the shape and the thickness, $d$, of the depth profile of the concentration of the charge carriers that are confined to the LAO/STO interface. In the present case, it originates from the photo-generated charge carriers at the surface of the STO (001) single crystal. The evolution of the spectra in Figs. 1(b)-1(d) reveals that the concentration of the photo-generated carriers increases during the UV illumination. This is evident since, as the UV-illumination progresses, the dip feature near 866 cm$^{-1}$ becomes more pronounced and the Berreman-mode is enhanced and slightly blue-shifted. Furthermore, Fig. 1(e) confirms that a Berreman-mode feature and thus a significant fraction of the photo-generated charge carriers persist for a long time after the UV-radiation has been turned off. For a quantitative analysis we have fitted the spectra using a layer-model with a conducting STO surface layer that has a graded profile of the carrier concentration and an anisotropic mobility of the carriers in the lateral and vertical directions with respect to the sample surface, $\mu_{ab}$ and $\mu_\perp$, respectively. The details of this model are discussed in section II and also in Ref. [49, 55]. The best fits shown by the solid lines reproduce the experimental data rather well. The obtained depth profiles of the carrier concentration are displayed in Fig. 1(f), the corresponding values of $N_s$, $d$, $\mu_{ab}$ and $\mu_\perp$, are listed in Table I.

In the first place, this analysis reveals a surprising similarity between the photo-generated charge carriers at the surface of STO (001) and the ones that are intrinsic to the LAO/STO (001) interface. The depth profile of the carrier concentration has a similar shape and thickness, $d$, and, after the extended UV-illumination, the values of $N_s$ and $\mu$ are also comparable. This suggests that despite of the supposedly different origin of the electrons at the surface of STO (001) and at the LAO/STO (001) interface, photo-generated oxygen vacancies in the former and a polar discontinuity and subsequent electron transfer between LAO and STO in the latter, there exists a similar confining potential. This may well be a consequence of the extremely large dielectric constant of STO (at low temperature) which leads to an efficient screening of defects and determines the length scale of the confining potential.

Furthermore, the analysis reveals that (i) the value of $N_s$ nearly doubles in magnitude during the UV illumination, (ii) $\mu$ increases slightly and (iii) a sizeable fraction of the charge carriers ($N_s$ in Table I is reduced by about 50%) persists on a time scale of hours after the UV-radiation has been switched off, as long as the temperature is maintained at 30 K.

The temperature dependence of the persistent part of the Berreman-mode feature, after the sample was illuminated for 45 minutes and then switched off, is shown in Fig. 2. It reveals that the Berreman mode and thus the persistent photo-doped carriers are vanishing as the temperature is raised well above $T^* = 105$ K. The inset shows that the amplitude of the dip feature, which is a measure of $N_s$, exhibits an order-parameter-like temperature dependence and vanishes between 110 and 130 K. The cubic-to-tetragonal transition and the subsequent structural domain formation in the near surface region of STO are indeed well-known to occur at a somewhat higher temperature than in the bulk, i.e. around 120-130 K [10, 58]. The data in Fig. 2 thus confirm that the structural domain boundaries are a prerequisite for the long-lived, photo-generated electrons to occur. Notably, this does not exclude the possibility that some of the oxygen vacancies may be trapped by other kinds of defects. Nevertheless, since they could not be detected in our IR-experiment, the resulting, photo-generated charge carriers seem to have a much smaller concentration (and/or mobility) than the ones due to the trapping of oxygen vacancies at the structural domain boundaries.

### III.B) Photo doping of STO (110)

Additional support for this conjecture has been obtained from a photo-doping experiment on a STO (110) single crystal for which the structural domain state can be varied in a controlled way and even a mono-domain state can be induced. We have previously shown, by studying the anisotropy of the IR-active phonon modes with ellipsometry [13], that the structural domain state of STO (110) can be strongly modified if the sample is cooled under a moderate uniaxial stress of only a few MPa. The latter has been applied by mounting the sample in a spring-loaded clamp. In this optical experiment the information about the domain orientation is obtained from the anisotropy of the oscillator strength of the so-called R-mode at 438 cm$^{-1}$. This mode only becomes IR-active below $T^* = 105$ K due to the doubling of the unit cell and a subsequent back-folding of the phonon modes from the zone boundary (the R-point) to the center of the Brillouin zone [59, 60]. The oscillator strength of this R-mode is proportional to the magnitude of the antiphase rotation of the TiO$_6$ octahedra and is maximal for the component perpendicular to the rotation axis of the oxygen octahedra (to the tetragonal axis) and zero for the parallel one. The intensity ratio of the R-mode along the [001] and the [1-10] directions is therefore a sensitive measure of the orientation of the structural domains. If it is zero along [001], then the sample is in a mono-domain state with the c-axis parallel to the [001] axis. If the ratio of the intensities along [001] and [1-10] is close to unity, then the sample is in a poly-domain state. In the former case, there are no domain boundaries that could trap the photo-generated oxygen vacancies and thus give rise to persistent charge carriers. In the latter case, there exists a fair amount of these structural domain boundaries. Their density is still expected to be quite a bit lower than in the STO (001) crystals for which the structural domains tend to be much smaller. The much larger size of the structural domains in pristine STO (110) crystals as compared to STO (001) was reported already in 1970 [12]. It seems that the intrinsic lateral anisotropy of the (110) surface provides a preferred orientation for the tetragonal axis to be along the [001] direction. For the (001) surface of STO there is no such surface anisotropy and the preferred orientation of the tetragonal axis is therefore rather determined by defects and related strain fields that break the local symmetry. Apparently, in STO (001) this leads to a random orientation of the structural domains and a rather small domain size with a high density of domain boundaries.

Figure 3 compares the time dependence of the DC resistance during and after UV-illumination for a STO (110) single crystal that has been prepared in different structural domain states. The evolution is shown in the pristine state with no uniaxial stress applied and thus with only few structural domains (red line), in the mono-domain state after cooling under uniaxial stress applied along the [1-10] direction (blue line), and in a poly-domain state with almost equal amounts of domains with the c-axis along [001] and [1-10] obtained by applying the uniaxial stress along [001] (green line). These measurements have been performed on the same STO (110) crystal using the same contacts such that the absolute values can be directly compared. Finally, the grey line shows the corresponding data of a STO (001) crystal in a poly-domain state with no external stress applied for which we expect that the density of the domain boundaries is significantly higher than the one in the poly-domain state of the STO (110) crystal.

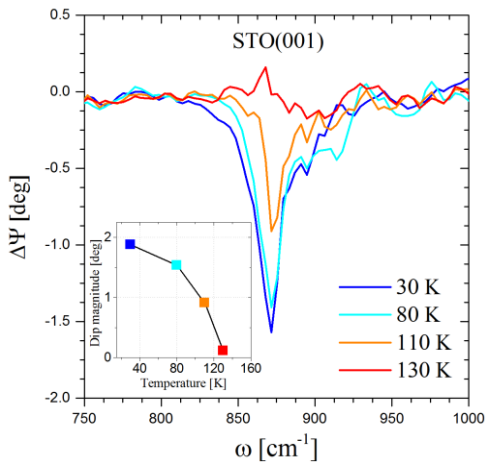

**Figure 2:** Difference spectra of the ellipsometric angle, $\Delta\Psi = \Psi(t \gg 0, T) - \Psi(t < 0, T)$, in the vicinity of the highest LO-mode of a STO (001) single crystal after and before UV-illumination at $T = 10$ K, as measured at different temperatures upon warming. The inset shows that temperature dependence of the magnitude of the dip around 865 cm$^{-1}$ that is an indicator for the sheet carrier density, $N_s$.

The most important observation in Fig. 3 concerns the time constant of the change of the DC resistance right after the UV light has been turned off. For the mono-domain sample the resistance rises instantaneously (within the time resolution of a few seconds) above the upper measurement limit of about 40 MΩ. For the pristine STO (110) sample with only few structural domain boundaries this increase of the DC resistance is already noticeably slower and it takes about 30 seconds until the upper resistance limit is reached. For the STO (110) crystal in the poly-domain state, after a small instantaneous increase, the DC resistance rises much more gradually and it takes around 7 minutes to reach the upper resistance limit. Last but not least, for the STO (001) crystal, after an initial increase, the DC resistance remains nearly constant at a value of R ≈ 2.25 MΩ (see also Fig. 1(a)). This highlights that the structural domain boundaries are playing an essential role in the photo-generation of the persistent charge carriers. A very likely scenario is that they act as trapping centres for the photo-generated oxygen vacancies (or photo-mobilized oxygen vacancies which migrate from the bulk to the surface) which give rise to the itinerant electrons. The pinning of the oxygen vacancies may be caused by the local strain and, as was mentioned already in the introduction, by a polar charging of the domain boundaries due to a so-called flexoelectric effect [21-23]. The formation of oxygen vacancy clusters may also play an important role since it can give rise to a further reduction of the Coulomb-interaction with the itinerant electrons and thus enhance the lifetime of the photo-generated charge carriers.

The comparison of the DC resistance data during the UV-illumination also reveals some interesting trends. It shows that the structural domain boundaries lead to an increase of the DC resistance. In the mono-domain state (blue line) the DC resistance of the STO (110) crystal is indeed quite a bit lower than the one in the poly-domain state (green line), in the pristine state it is intermediate (red line). At a first glance, this seems to be a surprising result since a higher density of the photo-generated carriers is expected if the oxygen vacancies are trapped by the domain boundaries and therefore have a longer life-time. Nevertheless, this trapping of the oxygen vacancies likely causes a spatially inhomogeneous distribution of the itinerant charge carriers such that the increase of the DC resistance can be understood in terms of a "bottle neck" effect from the carrier depleted regions. For the STO (110) crystal, unfortunately, we do not have complementary information about the evolution of the sheet carrier density from the IR-ellipsometry experiments. These have been performed in parallel to the DC resistance measurements, but did not show a clear spectroscopic signature of the confined electrons, e.g. no Berreman mode could be observed. This suggests that the concentration of the photo-generated charge carriers in STO (110) is more than one order of magnitude lower than the one in STO (001). It remains to be further explored whether this is a result of the lower concentration of the structural domain boundaries in STO (110) or rather caused by a difference in the electronic structure of the oxygen vacancies near the surface of STO (110) which may give rise to a stronger binding of the itinerant electrons.

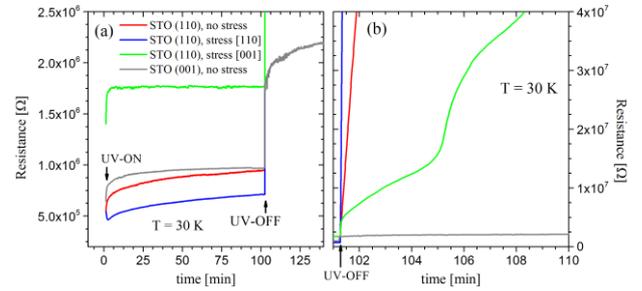

**Figure 3:** Photo-doping effect at 30 K on the DC resistance of a STO (110) single crystal in the weakly twinned pristine state (red line), in a mono-domain state with a weak stress applied along the [1-10] direction (blue line), and in a poly-domain state with the stress along [001] (green line). Also shown, for comparison is the corresponding curve of a heavily twinned STO (001) crystal (grey line). **(a)** Time dependence of the resistance during and right after the UV-illumination that was started at $t^{start}$ = 0 and stopped at $t^{stop}$ = 101 minutes, as indicated by the vertical arrows. **(b)** Magnification of the time-dependence right after the UV radiation has been switched off.

Another interesting trend in Fig. 3(a) concerns the time-dependence of the DC resistance immediately after the UV-light has been switched on. In the mono-domain state of STO (110) the DC resistance decreases rather rapidly within the first 30 seconds before it saturates and starts to increase much more gradually. In the poly-domain state of STO (110) and STO (001), instead, the DC resistance shows an initial increase which occurs on a similar time scale, i.e. within about 30 seconds. The decrease of the DC resistance in the mono-domain sample is a testimony that this phenomenon is not caused by a heating of the sample surface due to absorption of the UV-radiation. The initial decrease of the DC resistance rather seems to be governed by an increase of the carrier concentration which is determined by the rates of the photo-generation of the carriers and their recombination and trapping on defects, like the oxygen vacancies. Likewise, the corresponding increase of the DC resistivity in the poly-domain state can be understood in terms of the "bottle neck effect" from the carrier poor regions which occur when the oxygen vacancies agglomerate at the structural domain boundaries. The scattering of the itinerant carriers on these domain boundaries may provide an additional contribution to this increase of the DC resistance. The following, much more gradual increase of R seems to be determined by defects other than these domain boundaries, since it is also observed in the mono-domain state of STO (110).

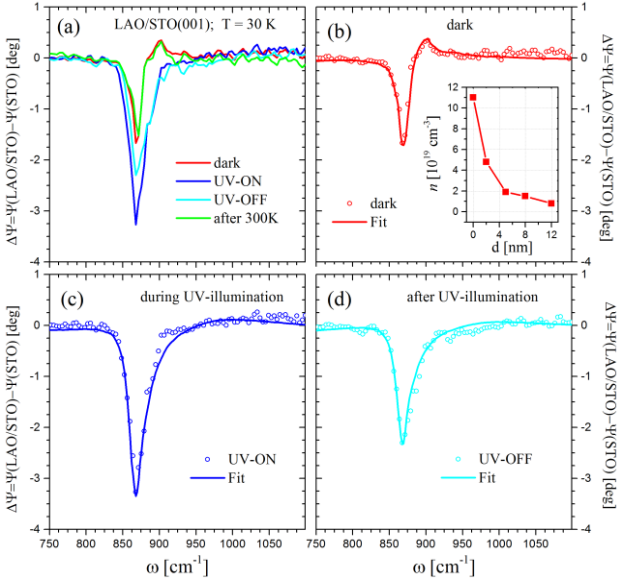

**Figure 4**: Difference spectra of the ellipsometric angle, ΔΨ = Ψ(LAO/STO) - Ψ(STO), of a LAO/STO (001) heterostructure with respect to a bare STO (001) substrate (without photo-doping) showing the Berreman mode in the vicinity of the highest LO phonon of STO and its variation upon photo-doping. **(a)** Comparison of the spectra at 30 K in the initial, dark state (red line), during UV-illumination (blue line), shortly after the UV radiation has been switched off (cyan line), and after the sample has been warmed to room temperature and cooled again (in the dark) to 30 K (green line). **(b)** Experimental data (red circles) and best fit (red line) to the data in the initial dark state at 30 K prior to the UV-illumination. The inset shows the obtained depth-profile of the carrier density. **(c)** Data (blue symbols) and best fit (blue line) of the spectrum obtained during UV illumination. **(d)** Data (cyan symbols) and best fit (cyan line) of the spectrum as obtained at 30 K shortly after the end of the UV illumination. The parameters obtained from the best fits in (b)-(d) are listed in Table II.

### III.C) Photo-doping of LAO/STO (001)

A pronounced photo-doping effect has also been observed in LAO/STO (001) heterostructures. Here it was previously shown that mobile electrons exist at the interface even without any UV-illumination [1, 2]. The origin of these intrinsic charge carriers is the subject of an ongoing discussion. The scenario of a so called "polar catastrophe" assumes that the electrons are transferred from the topmost layer of LAO to the one of STO as to avoid a divergence of the electric potential that arises from a discontinuity in the polarity of the atomic layer stacking at the LAO/STO (001) interface [1, 2]. Nevertheless, such a diverging electrical potential could also lead to the attraction of a certain amount of oxygen vacancies [61] or the intermixing of $La^{3+}$ and $Sr^{2+}$ or $Ti^{4+}$ and $Al^{3+}$ cations across the interface [28].

The response of these confined electrons at the LAO/STO (001) was previously studied with IR-ellipsometry in Ref. [49]. Fig. 4 shows the result of a photo-doping experiment on the same LAO/STO (001) heterostructure. The red line in Fig. 4(a) displays the IR-response in the vicinity of the Berreman-mode as obtained after cooling the sample to 30 K in dark conditions. The spectrum agrees well with the published data, the parameters obtained from the best fit are listed in Table II. The best fit shown by the solid line in Fig. 4(b) yields a sheet carrier density of about $4.2 \times 10^{13}$ cm$^{-2}$ and a depth distribution of the carrier concentration as shown in the inset of Fig. 4(b). Apparently, the concentration and the depth distribution of the charge carriers did not change significantly during several years.

On the other hand, the response of the itinerant charge carriers becomes much more pronounced after the same LAO/STO (001) heterostructure has been illuminated with UV-light. The blue line in Fig. 4(a) shows that the spectral features are strongly enhanced and the corresponding best fit in Fig. 4(c) yields a three-fold increase of the sheet carrier density as compared to the original state (see Table II). Furthermore, a surprisingly large fraction of these photo-generated charge carriers persists even after the UV-radiation has been switched off. This is shown by the cyan line in Fig. 4(a) and the best fit in Fig. 4(d) which indicates a sheet carrier density that is still twice as large as before the UV-illumination.

TABLE II. Parameters of the best fits to the ellipsometry data shown in Figs. 4(b)-(d) using the model of an isotropic electron gas as described in Ref. [47].

| LAO/STO (001) | $N_s$ [$10^{13}$ cm$^{-2}$] | $d$ [nm] | $\mu$ [cm$^2$/Vs] |
|---|---|---|---|
| Dark; (Fig. 4b) | 4.2±0.6 | 12±2 | 21±3 |
| UV-ON ; (Fig. 4c) | 15±0.5 | 19±2 | 20±3 |
| UV-OFF; (Fig. 4d) | 9.6±0.5 | 13±2 | 21±3 |

In sections III.A and III.B we have discussed the evidence that the persistent photo-generated charge carriers at the surface of STO (001) and STO (110) single crystals originate from oxygen vacancies that are trapped by the structural domain boundaries which appear below $T^*$. The data in Fig. 4 suggest that these oxygen vacancies can also give rise to a substantial enhancement of the sheet carrier density at the LAO/STO (001) interface.

Similar to the STO (001) surface (see Fig. 2), these photo-generated charge carriers are only persistent if the temperature remains below $T^*$, i.e. as long as the structural domain boundaries are present. The green line in Fig. 4(a) shows that the IR-response prior to the UV-illumination has been recovered after heating the sample to room temperature and cooling it again to 30 K in dark condition. On the other hand, the intrinsic sheet carrier density of the initial state is hardly affected by an annealing treatment that is performed in oxygen atmosphere at a temperature as high as 600 °C (data not shown). This suggests that the permanent charge carriers are either caused by a mechanism that is unrelated to oxygen vacancies, like the polar catastrophe, or that these oxygen vacancies are extremely stable (and fully saturated already in the pristine state). In the latter case, the polar discontinuity may still be responsible for the strong attraction of the oxygen vacancies to the LAO/STO (001) interface.

### III.D) Photo-doping and anisotropic free carrier response of LAO/STO (110)

The oxygen vacancy scenario has been further explored with LAO/STO (110) heterostructures for which, nominally, the interfacial layer stacking does not give rise to a polar discontinuity. DC transport measurements have still provided clear evidence for itinerant electrons with a concentration and mobility similar to the ones in LAO/STO (001) [30]. The ellipsometry spectra in Fig. 5(a) also reveal a clear Berreman-mode feature and thus confirm the existence of itinerant charge carriers at the LAO/STO (110) interface. In this measurement the sample has been shielded against UV radiation while it was cooled to 30 K to avoid photo-doping effects. The Berreman-mode exhibits a pronounced anisotropy between the [001] and [1-10] directions which suggests that the charge carrier mobility is strongly anisotropic. The best fit with a model that allows for an anisotropic in-plane mobility (solid lines) yields a sheet carrier density of $N_s = 3.1 \times 10^{13}$ cm$^{-2}$ (see Table III) similar to the one in LAO/STO (001) with $N_s = 4.2 \times 10^{13}$ cm$^{-2}$. The obtained mobility is about an order of magnitude higher along [001] than along [1-10]. A similar anisotropy was obtained in Ref. [62] from DC transport measurement on a LAO/STO (110) heterostructure that was also grown with an oxygen partial pressure of $10^{-4}$ mbar. This anisotropy furthermore agrees with band-structure calculations [63] which suggest that the lowest conduction band at the STO (110) surface is derived from the Ti orbitals with $d_{xz}$ and $d_{yz}$ symmetry which disperse much more strongly along [001] than along [1-10]. It is also consistent with previous measurements of the resonant x-ray linear dichroism which reveal that the $3d_{xz}$ and $d_{yz}$ orbitals of Ti are slightly lower in energy than the $3d_{xy}$ orbital [64].

Nevertheless, in the following we show that this anisotropy of the charge carrier mobility can be easily modified and even inverted, for example, if the sample is annealed slightly above room temperature, illuminated with UV radiation, or cooled under a weak uniaxial stress.

TABLE III. Parameters of the best fits to the ellipsometry data shown in Fig. 5 using the model of an anisotropic electron gas as described in Ref. [53].

| LAO/STO (110) | $N_s$ [$10^{13}$ cm$^{-2}$] | $d$ [nm] | $\mu_\perp$ [cm$^2$/Vs] | $\mu_{[1-10]}$ [cm$^2$/Vs] | $\mu_{[001]}$ [cm$^2$/Vs] |
|---|---|---|---|---|---|
| Pristine; (Fig. 5a) | 3.1±0.5 | 6±2 | 10±6 | 3±2 | 30±11 |
| Annealed at 120 °C; (Fig. 5b) | 3.1±0.5 | 6±2 | 10±6 | 34±11 | 3±2 |
| UV-ON; (Fig. 5d) | 8.0±0.9 | 9±2 | 15±4 | 15±3 | 15±3 |
| 1 year later; (Fig.5f) | 0.9±0.5 | 3±2 | 7±3 | 1±3 | 9±4 |

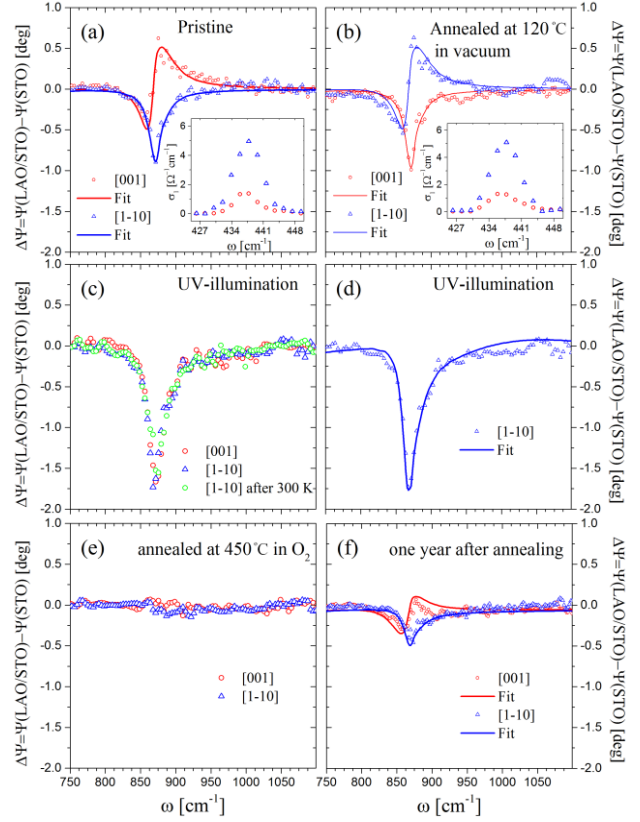

**Figure 5**: Difference spectra of the ellipsometric angle, $\Delta\Psi = \Psi(\text{LAO/STO}) - \Psi(\text{STO})$, of LAO/STO (110) with respect to a bare STO (110) substrate (without photo-doping) showing the Berreman mode in the vicinity of the highest LO phonon of STO and its variation upon UV-illumination and thermal annealing. **(a)** Spectra in the pristine state prior to any UV-illumination with the plane of incidence of the IR light along the [001] (red circles) and [1-10] direction (blue triangles). The best fits obtained with a model that allows for an in-plane anisotropy of the charge carrier mobility are shown by the corresponding solid lines. Inset: Spectra of the optical conductivity at the position of the so-called R-mode which has a much larger oscillator strength along [1-10] than along [001]. **(b)** Corresponding spectra as in (a) after the sample has been subject to a moderate thermal treatment for 1 hour at 120 °C and $10^{-6}$ mbar. The strong anisotropy of the Berreman mode is now inverted as compared to the pristine state in (a). **(c)** Spectra obtained shortly after a UV-illumination (for 45 minutes) measured along the [001] (red circles) and [1-10] directions (blue triangles). Shown by the green symbols is a spectrum after heating to 300 K and cooling again (in dark condition) to 10 K. **(d)** Comparison of the experimental data (blue triangles) and the best fit (blue line) of the spectrum along [1-10] shortly after the UV-illumination. **(e)** Ellipsometric spectra showing the absence of a Berreman mode after the sample has been annealed for 2 hours at 450 °C in flowing oxygen gas atmosphere. **(f)** Ellipsometric spectra of the Berreman mode that has reoccurred about 1 year after the annealing treatment in (e). The parameters obtained from the best fits in (a), (b), (d) and (f) are listed in Table III.

Figure 5(b) shows the Berreman-mode feature as measured on the same sample after it had been annealed for about 1 hour at 120 °C while it remained mounted in the cryostat in a vacuum of about $10^{-6}$ mbar. This gentle thermal treatment has been applied with the intention of removing residual organic solvents or other remnants from the surface of the LAO layer. For some of the LAO/STO (001) heterostructures it has been successfully used to obtain a moderate increase of the charge carrier mobility. For the LAO/STO (110) heterostructure there is a much more drastic effect on the Berreman-mode. The comparison of Figs. 5(a) and 5(b) shows that it leads to an inversion of the anisotropy of the Berreman mode as compared to the pristine state. Whereas the values of $N_s$ and $\mu_\perp$ are barely affected, the carrier mobility is now larger along [001] than along [1-10]. Note that the information about the crystallographic orientation of the sample is contained in the same ellipsometric spectra in terms of the anisotropy of the oscillator strength of the R-mode at 438 cm$^{-1}$ that is shown in inset of Figs. 5(a) and 5(b). An erroneous assignment of the crystallographic axis thus can be excluded.

This thermally-induced inversion of the anisotropy of the charge carrier mobility suggests that it is not only determined by the band-structure of the STO (110) surface (or interface). An important, additional role seems to be played by the oxygen vacancies. These can influence the anisotropy of the electronic response by forming extended clusters with a high aspect ratio and a preferred orientation and thus lead to anisotropic percolation and scattering effects. An alternative scenario is that the mobile defects (e.g. the oxygen vacancies) somehow induce an orbital reconstruction of the interfacial Ti levels and thus invert the $d_{xy}$ related and $d_{xz}$ and $d_{zy}$ related bands and the related anisotropy of the band dispersion. The latter scenario could be tested with XLD measurements of LAO/STO (110) during a similar thermal treatment.

Next we show that the UV-illumination provides another means to affect the charge carrier mobility and also to strongly increase the sheet carrier density. The spectra in Fig. 5(c) have been obtained after the LAO/STO (110) sample was illuminated with UV radiation for about 20 minutes at 30 K and switched off again. The spectra along the [001] and [1-10] directions are now almost identical within error bars, i.e. the charge carrier mobility is now almost isotropic. The best fit to the spectrum along [1-10] shown by the blue solid line in Fig. 5(d) reveals that the sheet carrier density is strongly increased to $N_s = 8\times10^{13}$ cm$^{-1}$ (see Table III). Notably, unlike in LAO/STO (001), these photo-doped charge carriers persists even after the sample has been warmed up to room temperature and subsequently cooled again to 30 K, see Fig. 5(c). This illustrates that the history of the UV illumination can strongly influence the concentration and even the anisotropy of the in-plane mobility of the charge carriers. The suppression of the anisotropy of the in-plane mobility might be a consequence of a rising Fermi-level, which leads to a partial occupation of the $d_{xy}$ band for which the anisotropy of the dispersion is opposite to the one of the $d_{xz}$ and $d_{yz}$ bands. An alternative explanation in terms of the oxygen vacancy clustering scenario is that the additional oxygen vacancies created by the UV-illumination are randomly oriented and require more time before they develop extended clusters. The relevant scenario could be identified with a series of repeated experiments that cover a longer time period.

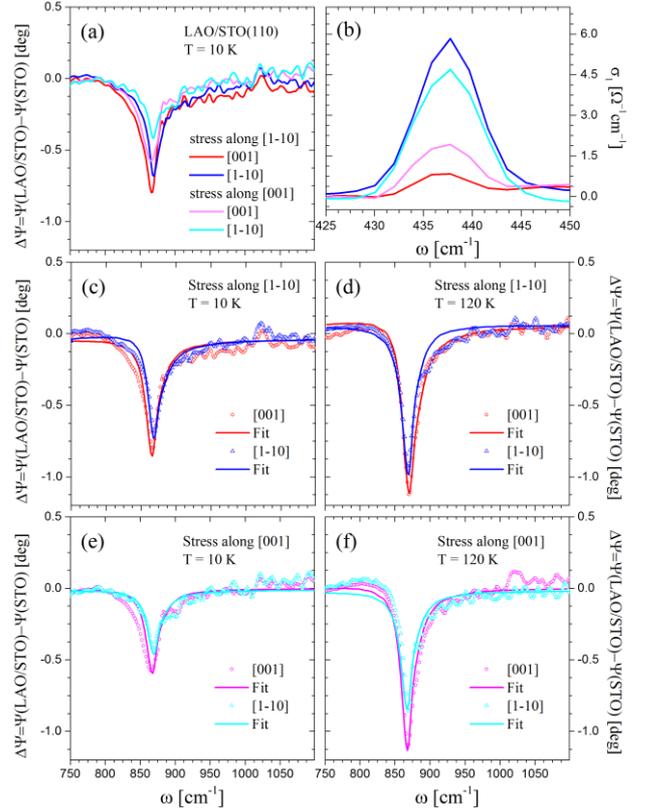

**Figure 6**: Effect of an external stress on the difference spectra of the ellipsometric angle, $\Delta\Psi = \Psi(\text{LAO/STO}) - \Psi(\text{STO})$, of a LAO/STO (110) heterostructure with respect to a STO (110) substrate. The measurements have been performed after the ones shown in Fig. 5(f). A weak uniaxial stress of about 2.3 MPa was applied along either the [1-10] or [001] direction by mounting the sample in a mechanical clamp and cooling it to 10 K. **(a)** Spectra showing the anisotropic response along the [1-10] and [001] directions with the stress applied along [1-10] (blue and red lines) or [001] (magenta and cyan lines). **(b)** Corresponding spectra of the R-mode which show the anisotropy of the oscillator strength along [1-10] and [001] from which the preferred orientation of the structural domains has been deduced. **(c)** Data (open symbols) and best fits (solid lines) for the $\Delta\Psi$ spectra at 10 K after cooling with uniaxial stress along [1-10]. **(d)** Corresponding data (symbols) and best fits (solid lines) obtained after warming the sample to 120 K. **(e)** Data (open symbols) and best fits (solid lines) for the $\Delta\Psi$ spectra at 10 K after cooling with the uniaxial stress along [001]. **(f)** Corresponding data (symbols) and best fits (solid lines) obtained after warming the sample to 120 K. The parameters obtained from the best fits in (c) - (f) are listed in Table IV.

We have instead performed an annealing treatment for two hours at 450 °C in one bar of oxygen gas to test whether this can influence these persistent, photo-induced charge carriers. The ellipsometric data in Fig. 5(e) confirm that this thermal treatment has entirely removed the Berreman-mode feature. This drastic annealing effect corroborates the conjecture that these charge carriers originate from oxygen vacancies that are trapped in the vicinity of the LAO/STO (110) interface. It is still a fairly surprising result since, as described in section **II.A.**, right after the PLD growth the sample had undergone an oxygen annealing treatment and the transport measurements showed a conducting state that did not change appreciably over time [30].

Further evidence in favor of the oxygen vacancy scenario is presented in Fig. 5(f) which shows that the Berreman-mode, and thus the mobile charge carriers at the LAO/STO (110) interface, reoccurred after the sample had been stored for about 1 year in a desiccator that is made from Plexiglas. We suspect that the related oxygen vacancies either have been photo-generated by some residual UV light or have slowly migrated from the bulk to the LAO/STO (110) interface. The best fit (solid line) shows that the carrier concentration is about 3 times lower than in the pristine state and, most notably, that the in-plane mobility exhibits the same kind of anisotropy as in the pristine state.

Finally, we show in Fig. 6 that a weak external stress applied along the [1-10] or [001] direction, which is known to reduce or enhance, respectively, the amount of structural domain boundaries in bare STO (110), provides yet another way of changing the sheet carrier density and also the anisotropy of the in-plane mobility. The experiments have been performed right after the one shown in Fig. 5(f), by mounting the sample in a clamp and cooling it to 10 K while great care was taken to shield it from UV radiation. The stress effect on the structural domain boundaries is shown in Fig. 6(b) in terms of the anisotropy of the oscillator strength of the R-mode. It shows that the stress along [1-10] ([001]) reduces (enhances) the oscillator strength of the R mode in the [001] spectrum and thus the number of domain boundaries (see Ref. [13]). Note that the stress along [1-10] does not give rise to a complete mono-domain state as in the case of bare STO (110) substrates [13]. This is likely due to defects and strain effects from the LAO layer which promote a poly-domain formation in the underlying STO (110) substrate.

First we discuss the stress effect on the sheet carrier density (see Table IV). The best fit to the spectra at 10 K with the stress along [1-10] (in the near mono-domain state) shown by the solid line in Fig. 6(c) yields a value of $N_s = 2 \times 10^{13}$ cm$^{-2}$ that is twice as large as the one in the initial state (see Fig. 5(f) and Table III). The corresponding fit in Fig. 6(e) shows that the sheet carrier density is reduced again to $N_s = 1 \times 10^{13}$ cm$^{-2}$ after the density of domain boundaries has been increased again by applying the stress along [001]. This is another confirmation that the structural domain boundaries have a strong influence on the itinerant charge carriers and act as trapping sites for the oxygen vacancies. The finding that the restoration of the structural domain boundaries is accompanied by a reduction of the sheet carrier density indicates that the domain boundaries are strongly trapping the oxygen vacancies and tend to localize the corresponding electrons more strongly than the other defects in the vicinity of the LAO/STO (110) interface.

TABLE IV. Parameters of the best fits to the ellipsometry data shown in Figs. 6(c)-(f).

| LAO/STO (110) | $N_s$ [$10^{13}$ cm$^{-2}$] | $d$ [nm] | $\mu_\perp$ [cm$^2$/Vs] | $\mu_{[1-10]}$ [cm$^2$/Vs] | $\mu_{[001]}$ [cm$^2$/Vs] |
|---|---|---|---|---|---|
| Stress [1-10]; 10 K; (Fig. 6c) | 2.0±0.6 | 5±3 | 5±3 | 25±8 | 10±5 |
| Stress [1-10]; 120 K; (Fig. 6d) | 1.0±0.5 | 11±4 | 15±6 | 60±22 | 27±5 |
| Stress [001]; 10 K; (Fig. 6e) | 1.0±1.0 | 3±2 | 7±3 | 9±4 | 1±3 |
| Stress [001]; 120 K; (Fig. 6f) | 1.0±0.5 | 12±4 | 11±7 | 48±19 | 30±11 |

We also observe a rather surprising stress effect on the charge carrier mobility. The comparison of the spectra in the initial state in Fig. 5(f) and in the state with the stress along [1-10] in Fig. 6(c) reveals that stress gives rise to yet another reversal of the charge carrier mobility.

Figure 6(e) shows that the anisotropy remains inverted after the number of structural domain boundaries has been increased again by applying the stress along [001]. Furthermore, the data in Figs. 6(d) and 6(f) show that this anisotropy of the mobility persists even when the temperature is raised above $T^*$. Another remarkable feature is the strong increase of the absolute value of the charge carrier mobility and also the thickness of the conducting layer toward higher temperature which has not been observed for any of the LAO/STO (001) heterostructures.

The above experiments demonstrate that the mobility of the itinerant charge carriers at the LAO/STO (110) interface and, especially, their in-plane anisotropy is strongly dependent on the history of the thermal treatment, the exposure to UV-radiation, and even the externally applied stress. The anisotropy of the in-plane mobility seems to be governed not only by intrinsic factors, like the anisotropic surface (or interface) band structure of STO (110) [61], but also by extrinsic factors due to extended defects, like the formation of oxygen vacancy clusters and their preferred orientation. These extended defects seem to undergo slow changes that involve time scales on the order of weeks or even months. A common theme of the above described experiments appears to be that the mobility is highest along [001] when the sample was not perturbed for a long time, whereas a strong reduction or even an inversion of the mobility occurred after the sample had been perturbed, either by a gentle annealing treatment, photo-doping, or a weak external stress. This suggests an important role of oxygen vacancy ordering and clustering phenomena and it underlines the versatility of the electronic properties of the LAO/STO (110) heterostructures.

## IV) DISCUSSION AND SUMMARY

With IR ellipsometry and DC transport measurements we investigated the mobile charge carriers which are induced by UV-illumination at the (001) and (110) surfaces of STO crystals and at the LAO/STO interfaces of corresponding heterostructures. We found that these photo-generated charge carriers, which most likely originate from oxygen vacancies, have rather similar properties as the ones that exist already in the dark state in the LAO/STO heterostructures, for which it is still discussed whether they originate from oxygen vacancies or from a so-called polar catastrophe. We have also obtained evidence that the trapping of the photo-generated oxygen vacancies at defects can strongly influence the life time of the related itinerant electrons. For the case of STO (001) and LAO/STO (001) the lifetime of the photo-generated charge carriers is very long at low temperature but decreases rapidly when the sample is heated above $T^*$. This highlights that the structural domain boundaries, which form below the antiferrodistortive transition at $T^*$, are playing an important role in trapping the oxygen vacancies. This conjecture is confirmed by the photo-doping experiments on STO (110) crystals for which the domain boundaries can be readily modified and even removed by applying a weak external stress. Evidence for an additional and even stronger trapping mechanism of the oxygen vacancies has been obtained for the LAO/STO (110) heterostructures for which the photo-generated charge carriers persist even after the UV light has been switched off and the sample has been heated to room temperature. Furthermore, these persistent photo-electrons, and even the ones that existed before the UV-illumination, can be removed by an annealing treatment at 450 °C in flowing oxygen gas. This is an indication for an oxygen vacancy trapping mechanism at the LAO/STO (110) interface that either does not exist at the LAO/STO (001) interface or is even much stronger and fully saturated, such that a certain concentration of interfacial oxygen vacancies exists irrespective of the thermal treatment. Finally, we found that the anisotropy of the charge carrier mobility in LAO/STO (110), which was previously assigned to surface band structure effects, can be strongly affected by a thermal treatment, UV illumination and even a weak external stress that modifies the structural domain boundaries. This highlights that the electronic properties of the LAO/STO (110) interface are extremely versatile and can be strongly affected by oxygen vacancy clustering and ordering phenomena. Notably, these ordering processes may involve very long time scales on the order of weeks or even months.

## ACKNOWLEDGMENT

The work at the University of Fribourg was supported by the Schweizerische Nationalfonds (SNF) through Grant No. 200020-153660. B.P.P.M. wishes to acknowledge support from the Marsden Fund of New Zealand. The work at Muni was financially supported by the MEYS of the Czech Republic. M.S., F.S., and G.H. acknowledge the support by the Spanish Government through Projects No. MAT2014-56063-C2-1-R, the Severo Ochoa SEV-2015-0496 grant and the Generalitat de Catalunya (2014SGR 734 project). J. Mannhart is acknowledged for providing the LAO/STO (001) sample and J. Foncuberta for scientific discussion.

†meghdad.yazdi@unifr.ch
‡christian.bernhard@unifr.ch